# Flexible Seamless 2-in-1 Design with Sample Size Adaptation

Runjia Li[a], Liwen Wu[b]*, Rachael Liu[b], Jianchang Lin[b]*

[a]*Department of Biostatistics, University of Pittsburgh, Pittsburgh, PA, United States;* [b]*Takeda Pharmaceuticals, Cambridge, MA, United States*

Corresponding author: Liwen Wu, liwen.wu@takeda.com; Jianchang Lin, Jianchang.Lin@takeda.com; Takeda Pharmaceuticals, Cambridge, MA 02139, USA

**Abstract:**

2-in-1 design (Chen et al. 2018) is becoming popular in oncology drug development, with the flexibility of using different endpoints at different decision time. Based on the observed interim data, sponsors choose either to seamlessly advance a small phase 2 trial to a full-scale confirmatory phase 3 trial with a pre-determined maximum sample size, or to remain in a phase 2 trial. This approach may increase efficiency in drug development but is rigid and requires a pre-specified fixed sample size. In this paper, we propose a flexible 2-in-1 design with sample size adaptation, while retains the advantage of allowing intermediate endpoint for interim decision. The proposed design reflects the needs of recent FDA's Project FrontRunner initiative to encourage using an earlier surrogate endpoint to potentially support accelerated approval with conversion to standard approval with long term endpoint from the same randomized study. Additionally, we identify the interim decision cut-off to allow conventional test procedure at the final analysis. Extensive simulation studies showed the proposed design require much smaller sample size and shorter timeline than the simple 2-in-1 design, while achieving similar power. A case study in multiple myeloma is used to demonstrate the benefits of the proposed design.

**Keywords:** Adaptive design, Seamless design, Sample size re-estimation, Surrogate endpoint

## 1. Introduction

Oncology drug development often demands innovations to accelerate patient access to new therapies. There have been numerous initiatives and adaptive designs proposed and under development, across industry and regulatory bodies, from early phase for dose optimization to registration for benefit-risk characterization. For instances, adaptive designs for dose finding, randomized cohorts for dose optimization, master protocol including multiple tumor types in phase I/II for the evaluation of antitumor activity, are actively discussed and implemented to expedite the development of oncology drugs and biologics (FDA 2022a). With the rapidly developing clinical landscape, a randomized phase 3 trial is often expected to demonstrate substantial clinical benefit compared with available therapies. Albeit the principle of accelerating patient access, 62% of the trials did not achieve results with statistical significance among 235 recently published phase III randomized clinical trials (RCTs) (Amiri-Kordestani and Fojo 2012). Thus, a separate randomized phase 2 proof-of-concept (POC) trial is still popular targeting on de-risking the phase III trial. From a more practical and efficient perspective, alternative approach has been proposed such as seamless Phase II/III design, which not only helps to avoid unnecessary significant investment but also stands from ethical consideration to protect patients. Alternative methods and programs have been built to balance the risk and efficiency. Accelerated approval is an FDA program to allow earlier approval of drugs, and expedite access to medicines based on the medicine's effect on a surrogate endpoint (FDA 2014, 2020), such as overall response rate (ORR) and minimal residual disease (MRD). Confirmatory studies are still required afterwards, and FDA will then decide whether to grant conventional approval or remove from market. Recently, an FDA initiative "Project FrontRunner" came to public attention to change the accelerated approval paradigm away from single-arm studies toward randomized clinical trials (RCTs) with an earlier endpoint to potentially support accelerated approval with conversion to standard approval from the same study (FDA 2022b; Liu 2022). Approval in earlier settings benefits patients' access to new therapies and has more potential to alter the course of the disease. Studying the drug in the early course of disease will provide better assessments of the treatment effects due to less confounding effects related to prior treatment or associated complications, and it also allows head-to-head comparisons to the established standard of care (FDA 2022b). With many advantages, the project brings some statistical challenges for clinical trial design at the same time. The most important one is how to utilize earlier surrogate endpoints to make informed interim decisions in connection to the long-term benefit, such as overall survival, with desirable operating characteristics under the Project FrontRunner.

Although some of the challenges could be handled by methodologies for seamless phase 2/3 designs, such question as how to use surrogate endpoint to make informed interim decision has not been fully answered. A seamless design combines trials/phases into a single trial and protocol (Li et al. 2020; Wu et al. 2022). It eliminates the white space between phases, and can serve more than one adaptive feature, such as sample size adaptions, population and dose selection. In an inferential seamless trial, data collected prior to the interim look also plays an inferential role after the interim look (FDA 2022b), and by combining data from both phase 2 and phase 3 stages at the final analysis, power could be preserved with reduced overall sample-size. However, most seamless designs focus on only one primary endpoint for both the interim and final analysis.

Chen et al. (2018) proposed a 2-in-1 adaptive phase 2/3 design to expedite oncology drug development, with the flexibility of using different endpoints at different decision times. At the interim look of 2-in-1 design, based on the observed data, researchers could choose either to seamlessly advance the small phase 2 trial to a full-scale confirmatory phase 3 trial with a pre-determined maximum number of subjects, or to continue with the small phase 2 trial. Under a mild correlation assumption, the overall type I error was proved to be controlled with conventional test statistic at the final analysis. Although the 2-in-1 design can expedite the drug development and reduce the expected sample size by the seamless phase 2/3 transition, it requires a pre-specified fixed sample size, which may lead to unnecessary long study duration and high cost due to the lack of sample size adaptation flexibility. Particularly, if the interim result is overwhelmingly positive, expanding to a phase 3 trial with prespecified large number of patients is counterintuitive. On the other side, it may also introduce the risk of failure given limited available data prior to the initiation of seamless phase 2/3 trial and the study may be powered over-optimistically.

One plausible solution is to incorporate sample size re-estimation (SSR) to the design to add flexibility to the sample size or event size based on available information at the interim analysis. Unblinded SSR is attractive because it requires smaller commitment upfront compared with the group sequential approach, and it can also be combined with other design features, such as dose or population selection. One main concern for unblinded SSR is the overall type I error inflation that has been demonstrated and discussed in many publications (Cui et al. 1999; Proschan and Hunsberger 1995). For seamless design with sample size adaption, there are many multiplicity adjustment techniques available in literature to control the overall type I error. Combination tests are often used for controlling the overall type I error across phase 2/3 stages, by combining the information from both stages in the final analysis with various combination functions (Li et al. 2020). Fisher's product test uses the product of two p-values for two stage as the final test statistic and compares with a critical value that is calculated from the corresponding quantile of $\chi^2$ distribution (Bauer and Kohne 1994). By applying different combination functions to the test statistics of two stages, Proschan and Hunsberger test, and CHW test obtain the weighted final test statistics and protect the overall type I error (Cui et al. 1999; Proschan and Hunsberger 1995). In the context of 2-in-1 designs, Cui et al. proposed a CHW 2-in-1 design to further improve the adaptive performance of 2-in-1 design by incorporating the CHW method (Cui et al. 2019). In this design, if the interim decision is advancing as a phase 3 trial, the sample size could be re-calculated based on conditional power. Nonetheless, CHW method assumes the same endpoint at the interim analysis as at the final analysis (i.e. the primary endpoint) to be used for adaptation (Bai and Deng 2019), thus it cannot offer the advantage of 2-in-1 design that allows different endpoints. Other multiplicity adjustment techniques may also be applied to this framework, for example, the conditional error approach (Proschan and Hunsberger 1995), but appropriate extension needs to be established to account for different endpoints at the interim and final analyses. Another type of methods is to control the overall type I error via adaptation rule. As proved by Chen et al. (2004), the overall type I error using conventional testing procedure is not inflated if the sample size is only increased when the unblinded interim result is promising. The interim result is considered promising when the conditional power, defined as the conditional probability of rejecting the null hypothesis at the final analysis given the interim results, is higher than 50%. Mehta and Pocock introduced a new adaptive algorithm to increase sample size by boosting the conditional power up to the targeted nominal power subject to a pre-specified upper limit (Mehta and Pocock 2011). The sample size is only increased when the conditional power falls in the promising zone, with the nominal power as the upper bound and a pre-determined $CP_{min}$ as the lower bound. However, there is no existing method to easily extend the concept of "promising zone" to 2-in-1 designs with different endpoints at the interim and final analyses.

In this paper, we propose a flexible seamless 2-in-1 design with sample size re-estimation to tackle these statistical considerations. In Section 2, we briefly review the 2-in-1 design proposed by Chen et al. (2018). Then, we introduce the proposed flexible seamless 2-in-1 design with sample size re-estimation and describe the methods to identify the decision threshold and control the overall type I error in Section 3. Simulation studies and results are presented to demonstrate the operating characteristics in Section 4. In section 5, an illustrative example is presented to demonstrate the time and resource-saving benefits of our proposed design in a real-world clinical trial setting. Lastly, we include discussions and conclusions in Section 6.

## 2. Simple 2-in-1 design

In the 2-in-1 design proposed by Chen et al. (2018) (denoted as S2in1 hereafter), using a short-term surrogate endpoint, an interim decision is made on whether to expand the small phase 2 proof-of-concept trial into a large phase 3 confirmatory trial or to remain as a phase 2 trial. If the observed interim result doesn't reach a pre-specified decision threshold, the trial continues as a phase 2 trial, with a planned smaller number of patients to be enrolled after the interim look. If the interim readout crosses the decision threshold, a pre-planned number of subjects will be enrolled in addition to the subjects already enrolled prior to the interim look, and then advances as a phase 3 trial (Cui et al. 2019).

Let $X$ denote the standardized test statistic for testing the treatment difference for the surrogate endpoint at the interim analysis, $Y$ denote the standardized test statistic for testing the treatment difference for the primary endpoint at the end of smaller phase 2 trial, and $Z$ denote the standardized test statistic for testing the treatment difference for the primary endpoint at the end of larger phase 3 trial. Without loss of generality, $X, Y, Z$ are assumed to follow standardized normal distributions under the null hypothesis that there's no treatment effect, with correlations $corr(X, Y) = \rho_{XY}$ and $corr(X, Z) = \rho_{XZ}$. Note that $Y$ and $Z$ can be the same endpoint based on different subjects, or different endpoints.

Suppose $n_1$ subjects are enrolled before the interim analysis. Based on these subjects, the interim statistic $X$ is compared with a pre-specified threshold $c$. If $X > c$, the trial is expanded to a larger phase 3 trial with additional $n_2$ subjects. At the end of the study, the test statistic $Z$ of the primary endpoint is calculated from a total of $n = n_1 + n_2$ subjects, to examine the treatment effect of the investigational arm comparing to the control. The null hypothesis is rejected if $Z >$ $Z_{1-\alpha}$, where $Z_{1-\alpha}$ is the $(1-\alpha)$th percentile of the standard normal distribution. If $X \leq c$, the trial continues as a phase 2 study and test the null hypothesis using $Y$ based on $n = n_Y$ subjects at the end of the study. Similarly, the null hypothesis would be rejected if $Y > Z_{1-\alpha}$. The overall type I error is expressed as

$$Type\ I\ error\ =\ \Pr(X > c, Z > Z_{1-\alpha}) + \Pr(X \leq c, Y > Z_{1-\alpha}).$$

Chen proved that the overall type I error is conservatively controlled below the nominal level $\alpha$ under a generally held assumption $\rho_{XY} \geq \rho_{XZ}$. It indicates that the correlation between the surrogate endpoint for adaptative decision at the interim analysis and the primary endpoint of the phase 2 study at the end of the study, is assumed to be larger or equal than that of the surrogate endpoint and the primary endpoint of the phase 3 study. Notice that the correlation between the test statistics depends on both the correlation between the two corresponding endpoints, if they are not the same, and the degree of overlap between the two underlying analysis populations. In practice, the assumption $\rho_{XY} \geq \rho_{XZ}$ is generally held, and the validity of this assumption has been discussed in detail by Chen et al. (2018). However, expanding the design to a fixed large sample size planned beforehand with limited information causes the lack of flexibility and risk on investment and patients benefit. Specifically, when the surrogate result is overwhelmingly positive, such decision is counterintuitive, and it may lead to unnecessarily long timeline and cost. On the other hand, if the prior assumption is too optimistic, such decision will often increase the trial failure rate. We performed a power study for S2in1 design under different scenarios (Figure S1 in Supplementary Materials).

### 3. Flexible 2-in-1 design with sample size adaptation

To develop a flexible seamless design that reflects the initiative of Project FrontRunner, we propose a flexible 2-in-1 design (F2in1) incorporating sample size re-estimation (SSR) at the interim analysis. Aiming to make the design more flexible, control the risk, shorten the study timeline, and save sample size, the proposed design starts with a relatively small upfront sample size and has additional subjects enrolled if the interim result is promising. Figure 1 shows the schema of the proposed design.

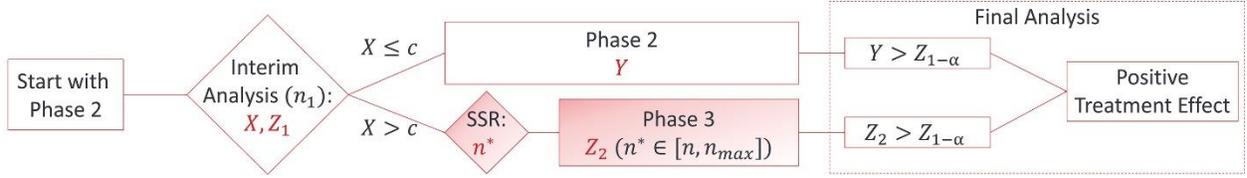

Figure 1. Schema of the proposed flexible seamless 2-in-1 design (F2in1) with sample size re-estimation (SSR).

## *3.1 Notations*

Let $O_T$ and $O_C$ be the responses if randomized to the treatment arm or control arm, respectively. We assume that they follow normal distributions with same variances, $O_T \sim N(\mu_T, \sigma)$, and $O_C \sim N(\mu_C, \sigma)$. RCTs usually focus on the treatment difference $\delta = \mu_T - \mu_C$. Here, we conduct one-sided test with a general null hypothesis $H_0: \delta = 0$ versus an alternative hypothesis $H_a: \delta > 0$. In this F2in1 design, following the similar notation as in S2in1 design, $X$ is the standardized test statistic of the surrogate endpoint at interim analysis, and $Y$ represents the endpoint at the end of phase 2 trial. $Z_1$ is the standardized test statistic of the primary endpoint for phase 3 trial observed at interim look, and $Z_2$ is the test statistic of the primary endpoint obtained at the end of phase 3 study. The study begins with $n_1$ subjects randomized 1:1 to the treatment and control groups. At the interim analysis, $X$ and $Z_1$ are observed based on these $n_1$ subjects. If $X \leq c$, where $c$ is a pre-specified decision threshold, the trial proceeds with additional $n_y - n_1$ subjects for a phase 2 study and test $Y$ based on $n_y$ subjects with the corresponding null hypothesis at the end of the study. If $X > c$, the trial proceeds with additional $n_2^*$ subjects for a phase 3 study and test $Z_2$ based on $n^* = n_1 + n_2^*$ with the corresponding null hypothesis at the end of the study. Let $n_2^*$ denote the adaptive sample size obtained at the interim based on the conditional power, $CP(z_1, n_2^*)$, defined as (Bauer and Koenig 2006; Mehta and Pocock 2011)

$$CP(z_1, n_2^*) = \Pr(Z_2 > Z_{1-\alpha}|z_1, n_2^*) = 1 - \Phi\left(\frac{Z_{1-\alpha}\sqrt{n} - z_1\sqrt{n_1}}{\sqrt{n_2^*}} - \frac{z_1\sqrt{n_2^*}}{\sqrt{n_1}}\right) = 1 - \beta, \quad (1)$$

where $1 - \beta$ is the nominal power. By Mehta and Pocock (2011), Gao and Mehta(2008), $n_2^*(z_1)$ from Equation (1) is

$$n_2^*(z_1) = \frac{n_1}{z_1^2}\left(\frac{Z_{1-\alpha}\sqrt{n} - z_1\sqrt{n_1}}{\sqrt{n_2}} + Z_{1-\beta}\right)^2, \tag{2}$$

where $Z_{1-\beta}$ is the $(1-\beta)^{\text{th}}$ percentile of the standard normal distribution. In Equation (2), $n_2^*(z_1)$ may be bigger or smaller than the original planned sample size $n_2$. In practice, we generally only consider sample size increasing and no decreasing, so the sample size after the interim analysis is adjusted only when $CP(z_1, n_2) < 1 - \beta$, which is equivalent to $Z_1 < W$, where $W = Z_{1-\alpha}\sqrt{t} + Z_{1-\beta}\sqrt{t(1-t)}$, and $t = n_1/n$ is the Fisher's information proportion at the interim analysis.

$$n_2^*(z_1) = \begin{cases} n_2 & \text{if } Z_1 \geq W \\ \frac{n_1}{z_1^2}\left(\frac{Z_{1-\alpha}\sqrt{n} - z_1\sqrt{n_1}}{\sqrt{n_2}} + Z_{1-\beta}\right)^2 & \text{if } Z_1 < W. \end{cases} \tag{3}$$

For practical considerations, trial sponsor usually sets up a maximum number of subjects to be enrolled in the study, $n_{max}$, such that $n^* = min(n_{max}, n_1 + n_2^*(z_1))$. This is also implemented in this proposed design. When $n_{max}$ is set to be infinity, there's no upper limit for the adjusted sample size $n^* = n_1 + n_2^*$. In this case, the conditional power is ensured to reach the nominal level, but an adjusted sample size larger than realistic may be unfeasible for patient recruitment. When $n_{max} = n$, and sample size $n^* = n = n_1 + n_2$ is not adjusted for any $z_1$. If so, the proposed design reduces to S2in1 with a fixed total number $n$ for the expanded phase 3 study, and the overall type I error can be controlled without any multiplicity adjustment under the same correction assumption specified by Chen et al. (2018). However, when $n_{max}$ is set too small, the conditional power may be inadequate. Therefore, $n_{max}$ should be pre-determined to balance the feasibility and the operating characteristics.

*3.2 Type I error control*

The proposed F2in1 design utilizes surrogate endpoint at the interim of phase 2 study with the sample size re-estimation feature to a reasonable shorten development timeline and ensure sufficient power at the end of the study. Multiplicity adjustment is a critical element during the sample size re-estimation procedure. The overall type I error is the probability of the null hypothesis being rejected either based on $Y$ at the end of phase 2 study or based on $Z_2$ after sample size re-estimation at the end of phase 3 study, i.e.,

$$Type\ I\ error = \Pr(Y > Z_{1-\alpha}, X \leq c) + \Pr(Z_2 > Z_{1-\alpha}, X > c). \tag{4}$$

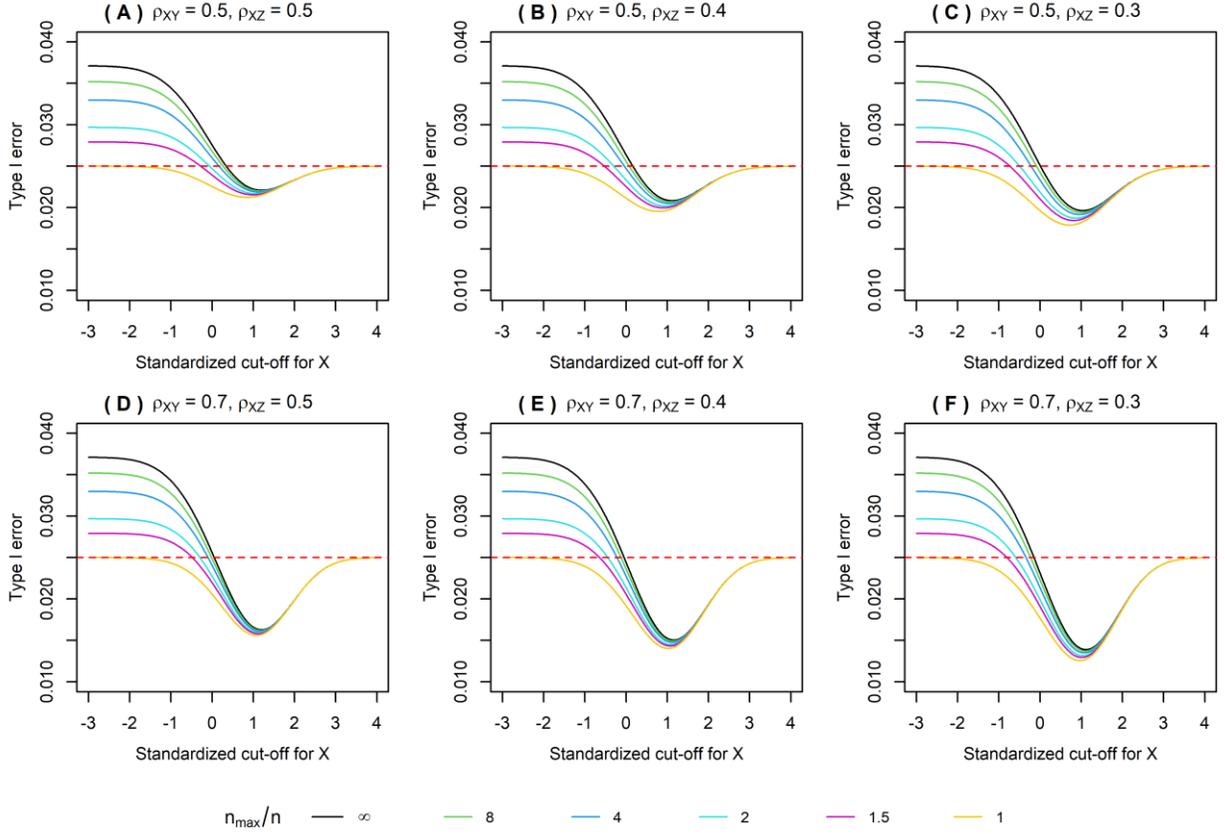

Figure 2. Overall type I error of the proposed design as a function of the standardized cut-off for the adaptation decision based on $X$ under different correlation compositions and prespecified maximum of adjusted sample size n. $\alpha = 0.025$ (one-sided), $1 - \beta = 0.9$.

The overall type I error could be inflated if the interim decision is to expand the study to a phase 3 trial and the sample size is increased, since the calculation for adjusted sample size and the final analysis at the end of phase 3 trial both depend on the primary endpoint at the interim look, $Z_1$ (Figure 2). When the interim decision threshold $c$ is small, it is likely to trigger sample size re-estimation at the interim analysis, and thus more type I error inflation. The higher the $n_{max}/n$, the more room for sample size adjustment. The largest inflation of type I error is observed when $n_{max}$ is set to be infinity, i.e. no restriction on how large the total sample size can get, and the interim decision boundary is set to be very low. However, this is not a practical case, since this design would always expand to a phase 3 study with unlimited resource, which is usually unrealistic. When $n_{max}/n$ is set to 1, the sample size is never adjusted regardless of the observed value of $Z_1$, and the overall type I error are controlled below the nominal level of 0.025 without any multiplicity adjustment. This result is consistent with S2in1, which can be seen as a special case of the proposed design. Similar to the findings in Chen et al. (2018), when the difference between $\rho_{XY} - \rho_{XZ}$ increases, the overall type I error tends to be smaller.

As shown in the Figure 2, the overall type I error is inflated over the nominal level when $n_{max} > n$ and $c$ is small. There is a cross point between the type I error curve and the horizonal line of the nominal level $\alpha$. On the left of the cross point, the type I error is inflated higher than the nominal level, while on the right, it is controlled under the nominal level. Thus, using $c$ that are larger than, i.e. on the right side of, the crossed point as the interim decision threshold, the type I error can be controlled under the nominal level while using conventional testing procedure at the final analysis. We denote this cross point as $C_{min}$,

$$C_{min}: \forall\ c > C_{min}, type\ I\ error\ < \alpha.$$

To control type I error in our F2in1 design without additional multiplicity adjustment, we propose to identify $C_{min}$, by solving the equation

$$Pr(Y > Z_{1-\alpha}, X \leq C_{min}) + Pr(Z_2 > Z_{1-\alpha}, X > C_{min}) = \alpha, \quad (5)$$

such that for any selected cut-off $c > C_{min}$, type I error will be controlled under the nominal level. $C_{min}$ can be obtained by either grid search in simulated empirical distribution under the null hypothesis, or by solving the Equation (5) analytically which offers optimal computational efficiency. We will focus on the latter in the rest of this section.

In Equation (5), the overall type I error consists of two parts. The first part is the probability of wrongly reject the null hypothesis at the end of phase 2 trial when the interim decision is not to expand. Because under null hypothesis, $\delta_X = \delta_Y = 0$, $X$ and $Y$ follow standard bivariate normal distribution with correlation $\rho_{xy}$, $Y|X = x \sim N(\rho_{XY}y, 1 - \rho_{XY}^2)$. The joint probability could be replaced by the product of conditional probability of rejection at phase 2 trial and density of a standardized normal distribution, i.e.

$$\Pr(Y > Z_{1-\alpha}, X \leq c)$$

$$= \int_{-\infty}^{c} \Pr(Y > Z_{1-\alpha}, X = x)\, dx$$

$$= \int_{-\infty}^{c} \Pr(Y > Z_{1-\alpha} | X = x)\, f_X(x)\, dx$$

$$= \int_{-\infty}^{c} \Phi\left(\frac{\rho_{XY} x - Z_{1-\alpha}}{\sqrt{1-\rho_{XY}^2}}\right) \phi(x)\, dx$$

(6)

The second part of the overall type I error is the probability of rejecting $H_0$ at the end of phase 3 trial. Since the statistic of the surrogate endpoint $X$ is based on the initial $n_1$ subjects, and the test statistic of the primary endpoint $Z_2$ at the final analysis is based on the initial $n_1$ and the additional $n_2^*$ subjects, it is challenging to directly write down the joint probability with the unknown correlation. Nevertheless, we can write the joint probability as the product of conditional joint probability of decision to phase 3 trial and rejection at the end of phase 3 trial given the observed phase 3 endpoint at the interim, and the density of $Z_1$, which is the standardize normal density, i.e.

$$\Pr(Z_2 > Z_{1-\alpha}, X > c) = \int_{-\infty}^{\infty} \Pr(Z_2 > Z_{1-\alpha}, X > c | Z_1 = z_1)\, f_{Z_1}(z_1)\, dz_1 \quad (7)$$

Given that $X$ and $Z_1$ are both statistics based on the first $n_1$ subjects and they are correlated, and $Z_2$ is based on the $n_1 + n_2^*$ subjects and follows the identical distribution with $Z_1$, following the independent increment assumption, $X$ and $Z_2$ are correlated only via $Z_1$. That is, $Z_2$ and $X$ are independent conditional on $Z_1 = z_1$ ($Z_2 \perp\!\!\!\perp X | Z_1 = z_1$). The conditional joint probability can then be further decomposed as the conditional probability of rejection at the end of phase 3 trial given $Z_1 = z_1$ and the conditional probability of expanding to phase 3 trial given $Z_1 = z_1$:

$$\Pr(Z_2 > Z_{1-\alpha}, X > c | Z_1 = z_1) = \Pr(Z_2 > Z_{1-\alpha} | Z_1 = z_1)\Pr(X > c | Z_1 = z_1) \quad (8)$$

Since the cumulative sample size is increased to $n^*$ only when $Z_1 < W$, we break the integration into two segments for easier calculation, $(-\infty, W)$ and $[W, \infty)$. The Equation (7) could then be expressed as

$$\int_{-\infty}^{W} \Pr(Z_2 > Z_{1-\alpha}|Z_1 = z_1) \Pr(X > c|Z_1 = z_1) f_{Z_1}(z_1) dz_1$$

$$+ \int_{W}^{\infty} \Pr(Z_2 > Z_{1-\alpha}|Z_1 = z_1) \Pr(X > c|Z_1 = z_1) f_{Z_1}(z_1) dz_1. \quad (9)$$

Notice that each part of the integrand is evaluable based on bi-variate normal distributions under the null hypothesis. To evaluate the semi-infinite integration, we applied numerical method to project it into a finite interval (details in Appendix). $C_{min}$ can then be solved directly from the Equation (10).

$$\alpha = \int_0^1 h_1(c) \frac{ds}{s^2} + \int_0^1 h_2(c) \frac{ds}{(1-s)^2} + \int_0^1 h_3(c) \frac{ds}{s^2},$$

(10)

where

$$h_1(c) = \Phi\left(\frac{\rho_{XY}\left(c - \frac{1-s}{s}\right) - Z_{1-\alpha}}{\sqrt{1-\rho_{XY}^2}}\right) \phi\left(c - \frac{1-s}{s}\right),$$

$$h_2(c) = \Phi\left(\frac{\left(W + \frac{s}{1-s}\right)\sqrt{t} - Z_{1-\alpha}}{\sqrt{1-t}}\right) \Phi\left(\frac{\rho_{XZ}\left(W + \frac{s}{1-s}\right) - c}{\sqrt{1-\rho_{XZ}^2}}\right) \phi\left(W + \frac{s}{1-s}\right),$$

$$h_3(c) =$$

$$\Phi\left(\frac{\left(W - \frac{1-s}{s}\right)\sqrt{t} - Z_{1-\alpha}\sqrt{\min\left(t + \frac{t}{\left(W - \frac{1-s}{s}\right)^2}\left(\frac{W}{\sqrt{t(1-t)}} - \left(W - \frac{1-s}{s}\right)\sqrt{\frac{t}{1-t}}\right)^2, \frac{n_{max}}{n}\right)}}{\min\left(\left|\frac{\sqrt{t}}{\left(W - \frac{1-s}{s}\right)}\left(\frac{W}{\sqrt{t(1-t)}} - \left(W - \frac{1-s}{s}\right)\sqrt{\frac{t}{1-t}}\right)\right|, \sqrt{\frac{n_{max}}{n} - t}\right)}\right) \Phi\left(\frac{\rho_{XZ}\left(W - \frac{1-s}{s}\right) - c}{\sqrt{1-\rho_{XZ}^2}}\right) \phi\left(W - \frac{1-s}{s}\right).$$

Table 1 presents the $C_{min}$ evaluated under various scenarios, with combinations of varying $\rho_{xy}, \rho_{xz}$, and $n_{max}$, to achieve designed level of power, $1 - \beta$, while controlling the overall type I error at the nominal level $\alpha$. The estimated values are obtained by solving Equation (5) directly and by grid search in simulated empirical null distributions. $C_{min}$ obtained from these two methods are basically numerically identical, but the analytical method we introduced in this section can obtain the desired value to fit specific design parameters much faster and more accurate than the simulation method, since the latter is more computationally intensive and always have random error. $C_{min}$ increases when the prespecified $n_{max}/n$ increases, which means as the upper limit for adjusted sample size set higher, there's more room for sample size adjustment and the type I error is more likely to inflate, so that we need a higher interim decision boundary to control the type I error. As the difference between $\rho_{XY}$ and $\rho_{XZ}$ increases, $C_{min}$ decreases, which is consistent with the trend of the overall type I error in Figure 2.

Table 1. Estimated $C_{min}$ in Equation (10) of the standardized cut-off for $X$. Empirical estimates is obtained by grid search on the empirical distribution by simulation (N=1,000,000). Theoretical estimate is obtained directly by solving Equation (10). Correlation between surrogate endpoint and primary endpoint in phase 2 trial is fixed as $\rho_{XY} = 0.7$, and correlation between surrogate endpoint and primary endpoint in phase 3 trial is $\rho_{XZ} = 0.5$.

| | | $C_{min}$ | | | | | | | | |
|---|---|---|---|---|---|---|---|---|---|---|
| | | $n_{max}/n = \infty$ | | $n_{max}/n = 8$ | | $n_{max}/n = 4$ | | $n_{max}/n = 2$ | | $n_{max}/n = 1.5$ | |
| $\rho_{XY}$ | $\rho_{XZ}$ | Theoretical | Simulation | Theoretical | Simulation | Theoretical | Simulation | Theoretical | Simulation | Theoretical | Simulation |
| 0.5 | 0.5 | 0.3442 | 0.3618 | 0.2665 | 0.2182 | 0.1683 | 0.1681 | -0.0603 | -0.0709 | -0.2737 | -0.2263 |
| 0.5 | 0.4 | 0.1634 | 0.1523 | 0.0755 | 0.0557 | -0.0396 | -0.0165 | -0.2963 | -0.2596 | -0.5240 | -0.5197 |
| 0.5 | 0.3 | 0.0100 | 0.0260 | -0.0862 | -0.0870 | -0.2148 | -0.2018 | -0.4946 | -0.4928 | -0.7309 | -0.7677 |
| 0.7 | 0.5 | 0.0516 | 0.0590 | -0.0140 | 0.0088 | -0.0989 | -0.0863 | -0.2947 | -0.2817 | -0.4764 | -0.4828 |
| 0.7 | 0.4 | -0.0481 | -0.0838 | -0.1223 | -0.1274 | -0.2209 | -0.2399 | -0.4423 | -0.4780 | -0.6412 | -0.6134 |
| 0.7 | 0.3 | -0.1475 | -0.1647 | -0.2296 | -0.2386 | -0.3411 | -0.3413 | -0.5877 | -0.6129 | -0.8005 | -0.7431 |

Other multiplicity adjustment techniques can also be considered to control overall type I error, for example, the CHW combination test (Bai and Deng 2019; FDA 2019). The combination test statistics $Z_{CHW} = Z_1\sqrt{n_1/n} + \tilde{Z}_2\sqrt{n_2/n}$, where $\tilde{Z}_2$ is the test statistic based on the data accumulated after the interim analysis and is independent from the information collected prior to the interim. $\sqrt{n_1/n}$ and $\sqrt{n_2/n}$ are two weights that need to be specified prior to the study at the design stage. The CHW test is often required by regulatory agencies to conservatively control the overall type I error when sample size re-estimation is planned in a study. However, a common criticism of this method is that, when sample size is increased, information from subjects enrolled after the interim analysis would be downweighed by the pre-specified weight based on initial designed sample sizes.

## 4. Simulation studies

### 4.1 Simulation setup

We performed extensive simulation studies to evaluate the operating characteristics of the proposed F2in1 design. The primary endpoints were set as time-to-event endpoints, with progression-free survival (PFS) for the phase 2 study and overall survival (OS) for the phase3 study. The early surrogate endpoint for the interim adaptation decision was set to be overall response rate (ORR). The total event size for the phase 2 trial was set as 118 to detect a hazard ratio 0.55 in PFS with 90% power and one-sided type I error 0.025. The designed event size for the cases of expanding to phase 3 trial was 180, which was calculated to detect a hazard ratio 0.617 in OS with 90% power and one-sided type I error 0.025. The interim analysis was planned to be conduct after 60 OS events are observed, corresponding to information fraction t=60/180=0.33. The interim analysis was planned early in the study so that the enrollment is less likely to be completed before the interim, and the adaptive decision could effectively mitigate the risk of overpowering the study with large number of subjects or committing to a lengthier development timeline. The maximum event size for expansion to phase 3 trial was capped at $m_{max}$ =330, that was calculated for detecting a 0.7 hazard ratio with 90% power, and it is used as a practical upper bound of the amount of the resources sponsor may invest in the study with the proposed design. Patients were equally randomized between the treatment and the control arms.

Under the null hypothesis, $HR_{OS} = HR_{PFS} = 1$, $ORR_c = ORR_t = 0.1$. Under the alternative hypothesis, various scenarios were explored. The first set of scenarios assume $HR_{OS} = HR_{PFS}$, which allowed us to compare the proposed method with the benchmark, S2in1 design, over a wider range of parameter space. The second set of scenarios assume $HR_{OS} \neq HR_{PFS}$, which were more realistic when the primary endpoints of the phase 2 and phase 3 studies are different by design. We further investigated the performance of F2in1 design, comparatively to the benchmark. For these scenarios, we assumed that the standardized test statistics for ORR ($X$) and PFS ($Y$) follow a standard bivariate normal distribution with correlation $\rho_{XY} = 0.7$, and similarly, $X$ and the standardized test statistic for OS ($Z_1$) at interim look to follow a standard bivariate normal distribution with correlation $\rho_{XZ} = 0.5$. Scenarios with varying ORRs, $\rho_{xz}$ and $m_{max}$ are also explored. In the simulation studies, we evaluated the operating characteristics of the proposed F2in1 design and compared it with the benchmark S2in1 design with total event size of the phase 3 study $m_{max}$ equals 180 or 330, since the final event size of the proposed design varies between this range. In the benchmark designs, the interim decision of whether to expand to the phase 3 study depends only on the interim observed value of $X$. Once the decision is to expand to phase 3, the trial continues until the fixed total event size, $m_{max}$, is reached. The empirical power, empirical type I error, expected number of events, study duration and number of patients were calculated overall and conditional on the interim decisions. Simulations were repeated for 100,000 iterations under the null scenario and 10,000 iterations under different alternative scenarios.

*4.2 Simulation results*

Table 2 shows the operating characteristics under the null hypothesis. Based on the design parameters set up for the simulation studies and the method described in Section 3.2, the threshold for the surrogate endpoint at the interim analysis to ensure type I error control was $C_{min} = -0.596$. When $c = C_{min}$, the probability of expanding to phase 3 trial is 73%, and the overall type I error is controlled at the nominal level 0.025. Notice that when designing such a trial, sponsor may expect trial to be more likely to stay as a phase 2 study under the null hypothesis, thus may not want to use $C_{min}$ as the cut-off for the interim analysis. To make clinically meaningful decisions, we set a threshold $c$ that is larger than $C_{min}$. As discussed in Section 3.2, the overall type I error would be controlled for all $c > C_{min}$, while still using the conventional test statistic at the final analysis. A clinical meaningful cut-off $c = 2.206$ was calculated based on $ORR_c = 0.1$ and $ORR_t = 0.25$, which means the trial would expand to a phase 3 study if at the interim analysis, the observed difference of ORR exceeds 0.15. In this scenario, more than 98% trial stay in phase 2 trial if there is no treatment effect in ORR. The decision threshold $c = 2.206$ were used in the following simulation of the alternative scenarios. The overall type I errors is controlled conservatively by using $c$ as compared with $C_{min}$, which is expected by the theoretical results derived in Section 3.2.

Table 2. Operating characteristics of the F2in1 design (number of Monte-Carlo iterations N=100,000) under null hypothesis, $HR_{OS} = HR_{PFS} = 1$, ORRs of control and treatment groups $ORR_c = ORR_t = 0.1$. Number of events before interim analysis $m_1 = 60$, number of events at final analysis for phase 3 trial in Chen's 2-in-1 design (S2in1) could be $m_1 + m_2 = 180$ or $m_{max} = 330$. Correlation between surrogate endpoint and primary endpoint in phase 2 trial is fixed as $\rho_{XY} = 0.7$, and correlation between surrogate endpoint and primary endpoint in phase 3 trial is $\rho_{XZ} = 0.5$. Cut-off c=2.206 of the intermediate test statistics is the mean statistics when $ORR_c = 0.1$, and $ORR_t = 0.25$, c = −0.596 is the solution of $C_{min}$ calculated from Equation (1).

| Cut-off c | Interim outcome | Probability of interim outcome | Type I error conditional on interim outcome | | |
|---|---|---|---|---|---|
| | | | S2in1 | | F2in1 |
| | | | $m_1 + m_2$ | $m_{max}$ | |
| -0.596 | Overall | - | 0.0227 | 0.0220 | 0.0256 |
| | $X \leq c$: Phase 2 | 27.56% | 0.0001 | 0.0001 | 0.0001 |
| | $X > c$: Phase3 | 72.44% | 0.0314 | 0.0304 | 0.0354 |
| 2.206 | Overall | - | 0.0204 | 0.0201 | 0.0204 |
| | $X \leq c$: Phase 2 | 98.67% | 0.0194 | 0.0194 | 0.0194 |
| | $X > c$: Phase3 | 1.33% | 0.0991 | 0.0736 | 0.0938 |

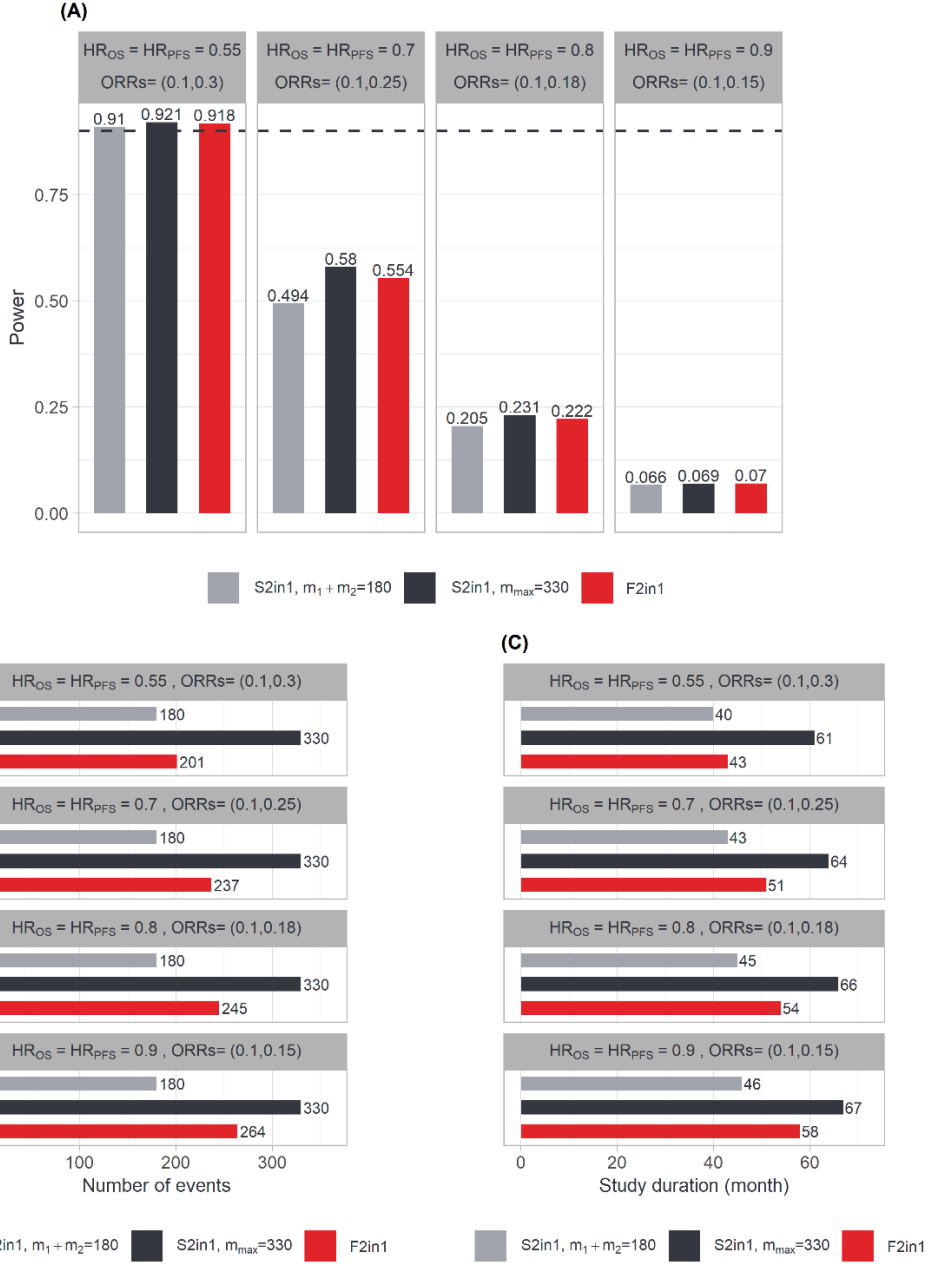

Figure 3. Average overall power (A), event size (B) and study duration (C) of the F2in1 design (number of Monte-Carlo iterations N=10,000) under alternative scenarios $HR_{OS}=HR_{PFS} \in \{0.55, 0.7, 0.8, 0.9\}$, ORRs=$(ORR_c, ORR_t) \in \{(0.1,0.3), (0.1,0.25), (0.1,0.18), (0.1,0.15)\}$, $\rho_{XY}=0.7$, $\rho_{XZ}=0.5$. Number of events before interim analysis $m_1 = 60$, number of events at final analysis for phase 3 in Chen's 2-in-1 design (S2in1) could be $m_1 + m_2 = 180$ or $m_{max} = 330$. Horizontal dash line represents designed power 0.9.

Table 3. Operating characteristics of the F2in1 design (number of Monte-Carlo iterations N=10,000) under alternative hypothesis with different true $HR_{OS}=HR_{PFS}$, ORRs (ORR of control and treatment groups). Number of events before interim analysis $m_1 = 60$, number of events at final analysis for phase 3 trial in Chen's 2-in-1 design (S2in1) could be $m_1 + m_2 = 180$ or $m_{max} = 330$. Power of using CHW statistics for the flexible 2-in-1 design with sample size re-estimation (F2in1) was also calculated for phase 3 trial (F2in1 w/ CHW). Correlation between surrogate endpoint and primary endpoint in phase 2 trial is fixed as $\rho_{XY} = 0.7$, and correlation between surrogate endpoint and primary endpoint in phase 3 trial is $\rho_{XZ} = 0.5$.

| $HR_{OS}$, $HR_{PFS}$, ORRs | Interim outcome | Probability of interim outcome | Power conditional on interim outcome | | | | Expected event size | | | Expected study duration | | |
|---|---|---|---|---|---|---|---|---|---|---|---|---|
| | | | S2in1 | | F2in1 | F2in1 w/ CHW | S2in1 | | F2in1 | S2in1 | | F2in1 |
| | | | $m_1+m_2$ | $m_{max}$ | | | $m_1+m_2$ | $m_{max}$ | | $m_1+m_2$ | $m_{max}$ | |
| 0.55, 0.55, (0.1,0.3) | Overall | - | 0.911 | 0.921 | 0.919 | - | 163 | 273 | 179 | 37 | 52 | 40 |
| | X<=c: Phase 2 | 26.7% | 0.703 | 0.703 | 0.703 | - | 118 | 118 | 118 | 29 | 29 | 29 |
| | X>c: Phase3 | 73.3% | 0.986 | 1.000 | 0.997 | 0.997 | 180 | 330 | 201 | 40 | 61 | 43 |
| 0.7, 0.7, (0.1,0.25) | Overall | - | 0.494 | 0.580 | 0.554 | - | 149 | 224 | 178 | 37 | 48 | 41 |
| | X<=c: Phase 2 | 50.1% | 0.241 | 0.241 | 0.241 | - | 118 | 118 | 118 | 31 | 31 | 31 |
| | X>c: Phase3 | 50.0% | 0.747 | 0.919 | 0.867 | 0.862 | 180 | 330 | 237 | 43 | 64 | 51 |
| 0.8, 0.8, (0.1,0.18) | Overall | - | 0.205 | 0.231 | 0.222 | - | 129 | 154 | 140 | 35 | 39 | 37 |
| | X<=c: Phase 2 | 82.9% | 0.145 | 0.145 | 0.145 | - | 118 | 118 | 118 | 33 | 33 | 33 |
| | X>c: Phase3 | 17.1% | 0.493 | 0.649 | 0.595 | 0.590 | 180 | 330 | 245 | 45 | 66 | 54 |
| 0.9, 0.9, (0.1,0.15) | Overall | - | 0.066 | 0.069 | 0.070 | - | 123 | 137 | 131 | 35 | 37 | 36 |
| | X<=c: Phase 2 | 91.1% | 0.050 | 0.050 | 0.050 | - | 118 | 118 | 118 | 34 | 34 | 34 |
| | X>c: Phase3 | 8.9% | 0.236 | 0.265 | 0.276 | 0.268 | 180 | 330 | 264 | 46 | 67 | 58 |

Figure 3 and Table 3 presents the overall power, expected event size and study duration, under various alternative hypothesis. The interim analysis was based on 120 patients after approximately 20 months after trial starts. The phase 2 trial ended with, on average, 200-220 patients after 29-34 months. Compared with S2in1 design with event size $m_{max} = 330$, the proposed F2in1 design shows similar overall power, but much smaller expected event size and much shorter study duration. For example, in the scenario of $HR_{OS} = HR_{PFS} = 0.55$ and $ORR_t = 0.3$, while the power of F2in1 design and S2in1 design are 0.919 and 0.921, respectively, the F2in1 design require 94 less events and save 1 year for the study timeline. The proposed design is most advantageous when the interim decision is to expand to a phase 3 trial. Conditional on advancing to a phase 3 study after the interim analysis, F2in1 requires much smaller event size and much shorter duration than S2in1 with similar power. In the same scenario highlighted above, F2in1 design requires 129 less events and 18-month shorter study duration than S2in1 design. Compared with the S2in1 design with event size $m = 180$, the proposed F2in1 design has much higher power, but just slightly larger event size and slightly longer study duration. The conditional phase 3 power using CHW test statistics is similar as F2in1 design, but it is calculated using weights, which adds complexity and may violate the principal that "all patients are equal" in clinical trials.

Table 4. Operating characteristics of the F2in1 design (number of Monte-Carlo iterations N=10,000) under alternative hypothesis with different true $HR_{OS} \neq HR_{PFS} = 0.55$, ORRs (ORR of control and treatment groups) = (0.1,0.3). Number of events before interim analysis $m_1 = 60$, number of events at final analysis for phase 3 trial in Chen's 2-in-1 design (S2in1) could be $m_1 + m_2 = 180$ or $m_{max} = 330$. Power of using CHW statistics for the flexible 2-in-1 design with sample size re-estimation (F2in1) was also calculated for phase 3 trial (F2in1 w/ CHW). Correlation between surrogate endpoint and primary endpoint in phase 2 trial is fixed as $\rho_{XY} = 0.7$, and correlation between surrogate endpoint and primary endpoint in phase 3 trial is $\rho_{XZ} = 0.5$.

| $HR_{OS}$ | Interim outcome | Probability of interim outcome | Power conditional on interim outcome | | | | Expected event size | | | Expected study duration | | |
|---|---|---|---|---|---|---|---|---|---|---|---|---|
| | | | S2in1 | | F2in1 | F2in1 w/ CHW | S2in1 | | F2in1 | S2in1 | | F2in1 |
| | | | $m_1 + m_2$ | $m_{max}$ | | | $m_1 + m_2$ | $m_{max}$ | | $m_1 + m_2$ | $m_{max}$ | |
| 0.7 | Overall | - | 0.709 | 0.852 | 0.816 | - | 163 | 272 | 213 | 40 | 55 | 47 |
| | X<=c: Phase 2 | 27.3% | 0.702 | 0.702 | 0.702 | - | 118 | 118 | 118 | 31 | 31 | 31 |
| | X>c: Phase3 | 72.7% | 0.711 | 0.908 | 0.858 | 0.852 | 180 | 330 | 248 | 43 | 64 | 53 |
| 0.8 | Overall | - | 0.451 | 0.6 | 0.572 | - | 164 | 274 | 237 | 42 | 57 | 52 |
| | X<=c: Phase 2 | 26.5% | 0.695 | 0.695 | 0.695 | - | 118 | 118 | 118 | 33 | 33 | 33 |
| | X>c: Phase3 | 73.5% | 0.364 | 0.566 | 0.528 | 0.508 | 180 | 330 | 280 | 45 | 66 | 59 |
| 0.9 | Overall | NA | 0.281 | 0.321 | 0.32 | | 163 | 274 | 253 | 43 | 59 | 56 |
| | X<=c: Phase 2 | 26.6% | 0.705 | 0.705 | 0.705 | | 118 | 118 | 118 | 34 | 34 | 34 |
| | X>c: Phase3 | 73.4% | 0.126 | 0.181 | 0.18 | 0.166 | 180 | 330 | 302 | 46 | 67 | 64 |

Scenarios with different $HR_{OS}$ and $HR_{PFS}$ were also simulated, and the results are shown in Table 4. When $HR_{OS}$ is not same as $HR_{PFS}$ and the overall treatment efficacy is optimal, we still observe similar advantage of F2in1 over S2in1 of being able to reach the similar power as requiring much smaller events, and shorter timeline. As $HR_{OS}$ increase, the overall power of both F2in1 and S2in1 decreases, as well as the power conditional on expanding to phase 3 trials. The power loss is mainly due to small effect size of the primary endpoint in the phase 3 study, but also caused by the inconsistency between surrogate and the primary endpoints, which leads to a higher probability of making incorrect interim decisions. For example, when $HR_{OS}$=0.9, $HR_{PFS} = 0.55$ and ORR difference is 0.2, the probability of expanding to phase 3 trial is 73.4%, resulting in the lower conditional probability of rejection at final analysis since the true effect size is small ($HR_{OS}$=0.9). This setting represents a potential scenario where the treatment largely improves the ORR and PFS, but not for OS. Under such scenario, 2-in-1 design may not perform well. Similar to all adaptive designs, careful planning and calibration is needed to ensure the performance of the proposed design is desirable.

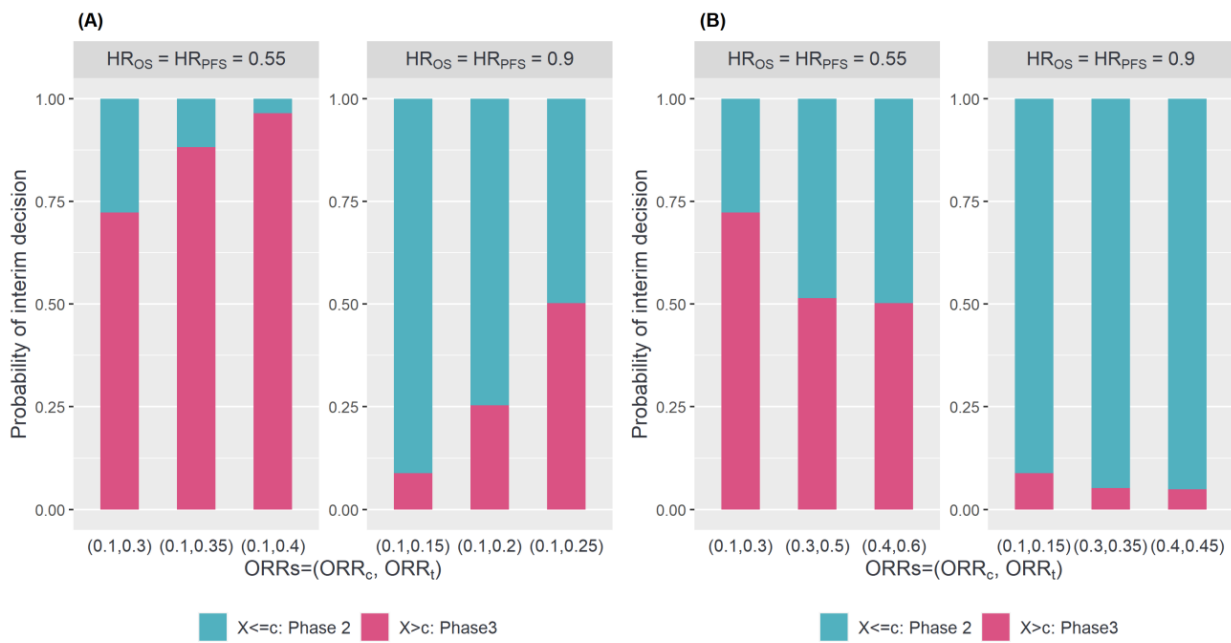

Figure 4. The probabilities of interim decisions given different ORRs by (A) Various level of ORR difference; (B) Varying ORR variance, over number of Monte-Carlo iterations N=10,000.

To further investigate the effect of varying ORRs on the interim decisions, Figure 4 shows the probability of interim decisions with varied ORR differences and variances while $HR_{OS}$ and $HR_{PFS}$ are fixed. As depicted in Figure 4 (A), when the ORR difference between the treatment and control groups increases, the probability of expanding to a phase 3 trial increases. When the ORR difference was fixed and the variance was increased, the probability of going to phase 3 trial decreases, indicating the probability of making right decision is lower when variance is large. Scenarios with different $\rho_{xz}$ and $m_{max}$ were also simulated, and the results provide similar trends and conclusions as above (Table S1 and Table S2 in Supplementary Materials).

## 5. Case study

In this section, we will present an illustrative example to demonstrate the time and resource-efficient benefits of our proposed design in a real-world clinical trial setting. Multiple myeloma is an uncommon type of blood cancer that affects patients' bone marrow. It is an uncurable disease, although patients have access to several recent development treatment options including proteasome inhibitors, immunomodulating drugs, and stem cell transplant therapy, that can improve the symptoms and prognostics. A new investigational treatment combination is under development for patients with relapsed and refractory multiple myeloma (RRMM). Early phase clinical studies have shown very promising safety profile and exciting treatment effects in terms on high overall response rate (ORR) for the combination therapy. Thus, the sponsor wants to explore the option of a seamless phase 2/3 design for late phase development to speed up in the rapidly evolving and highly competitive landscape.

To expedite the development while also remain fully flexible in terms of the overall development strategy, the sponsor decides to adopt a similar framework of the 2-in-1 design[9]. At a planned interim analysis, if the investigational therapy demonstrates promising treatment effect compared with the standard-of-care control arm, the trial will expand to a phase 3 stage to enroll more patients and perform final analysis with total information collected from both stages at the end of the trial. Nevertheless, if the investigational therapy doesn't meet the expansion criteria, the trial remains in phase 2 development with a fixed smaller sample size, and it allows the sponsor more time to collect evidence of the clinical benefits of the investigational therapy and to determine whether a large-scale confirmatory trial is warranted or not in the future. The primary endpoint of the phase 3 study is set to be overall survival (OS), and that of the phase 2 study is set to be progression free survival (PFS). Both are common endpoints to demonstrate treatment efficacy in late phase clinical development for RRMM. The interim analysis is planned early in the study, so an earlier intermediate endpoint and the key secondary endpoint of the study, ORR, is used to make more informed decision, given both PFS and OS may not be mature enough by the time. In RRMM studies, superior ORR has been proved to be a good proxy of survival benefit (Teng et al. 2018).

To ensure sufficient power at the end of the study, a sample size re-estimation procedure is also planned at the interim analysis, based on the phase 3 primary endpoint, OS. Depending on the interim observed treatment effects, the total number of OS events is expected to be between 180 and 330, which is required to detect an optimistic and moderate hazard ratio of 0.617 and 0.7 with 90% power, respectively. The total number of events expected at the end of the phase 2 study, if the interim decision is to remain in phase 2 development, is 118. It is calculated by powering a PFS hazard ratio of 0.55 at 90% level. The interim analysis is triggered after 60 OS events have been observed, corresponding to 0.33 information fraction, and the interim decision rule is following:

- if the testing statistic for ORR $< =2.206$, then continue the phase 2 trial and conduct the final analysis with additional 58 events;

- if the testing statistic for ORR $> 2.206$ then advance to phase 3 trial and conduct the final analysis with additional 120-270 events, and the exact number depends on the observed primary endpoint observed at interim analysis.

The threshold of the key secondary endpoint ORR is decided based on a clinically meaningful difference, 0.15 (0.1 in control arm and 0.25 in treatment arm), and is calibrated by simulation. Based on budget, enrollment, and other operational constraints, the trial is designed with maximum of 180 patients if it remains as a phase 2 study, and maximum of 500 patients if it expands to a phase 3 study.

The trial starts with the phase 2 stage and equally allocates $n_1=120$ patients between the investigational combination therapy and the control arm. The interim analysis is triggered after observing 60 OS event, and it is 21 months after study starts. The interim observed ORR difference is 0.227, and the standardized testing statistic is 3.157, so the interim decision is expanding to phase 3 trial. The observed hazard ratio of OS at the interim look is 0.664, and the conditional power is 81.3% (<90%). Following the event size re-estimation step discussed in Section 3.1 the event size after the interim analysis is increased to 167. After observing 227 events in total after 52 months, the final observed hazard ratio of OS is 0.749, with log rank test statistics 2.173 (p-value=0.015). The trial concluded with significant evidence of treatment effect of the investigational treatment over the control. If S2in1 design were used, the event size would be increased to the pre-specified maximum value 330, even with a such promising treatment efficacy demonstrated at the interim analysis.

The case study shows that the proposed F2-in-1 design substantially shorten the study timeline, save event size, and make the study more efficient. In this case, only 227 events are needed for the final analysis, 103 less than the simple 2-in-1 design. Compared with 64 months of study duration with the simple 2-in-1 design, the proposed design expects to trigger final analysis at 52 months. It makes big impact by allowing earlier access to new reliable treatment for the patients with serious conditions. By allowing conventional testing procedure at the final analysis, the proposed design is easy to implement. In summary, the F2in1 adds more flexibility to the simple 2in 1 design, demonstrates substantial time and resource saving properties, and is also simple and straightforward in practice.

## 6. Discussion

It became more common to accelerate the late phase drug development with limited internal data and bring medicine faster to patients. While directing to phase 3 trial can expedite the study, it brings potentially high risk and uncertainty with the limited data. The traditional learn-and-confirm study, starting phase 3 study after phase 2, requires larger sample sizes and may lead to longer trial duration to evaluate for the primary endpoint. Additionally, the decision making to transit from phase 2 to a separate phase 3 study is also vulnerable to trial-to-trial variability. As such, many regulatory initiatives have been promoted to expedite the drug development pathway, e.g. US FDA's Project FrontRunner, which encourage using an earlier surrogate endpoint to support potential accelerated approval with confirmational approval with long term endpoint from the same randomized study. In this paper, we proposed a flexible seamless 2-in-1 design with sample size re-estimation. It allows a seamless phase 2/3 transition based on a surrogate endpoint at the interim analysis, as well as sample size adjustment based on the unblinded primary endpoint results. While maintaining similar operating characteristics, the developed method is simpler but more flexible and efficient than other existing designs. Type I error is controlled with a proper cut-off $c > C_{min}$ for the surrogate endpoint, where the threshold $C_{min}$ can be solved numerically and accurately in the design stage. Compared with the S2in1 design, sample size re-estimation at the interim analysis can maximum the use of the observed data and make prompt adjustment to ensure sufficient power at the end of the study. It requires a smaller number of events and shorter study duration when the interim result is more promising, which is more reasonable and intuitive in practice. Compared with CHW 2-in-1 design (Cui et al. 2019), F2in1 design can save sample size while retains the advantage of allowing different surrogate endpoint for interim decision, which reflects the initiatives of Accelerated Approval Program and Project FrontRunner. Additionally, F2in1 design works well with conventional test statistics directly, and avoids the controversial and complicated weighting method, which has been suspected to violate the one-person-one-vote policy (Bai and Deng 2019; Fleming 2006).

There are additional statistical and regulatory considerations need to be taken into account to apply the flexible 2-in-1 design. Same as the simple 2-in-1 design, the correlation between the surrogate endpoint for adaptation decision and the primary endpoint at the end of phase 2 study, is assumed to be larger or equal than that of the surrogate endpoint and the primary endpoint of the phase 3 study, which is generally held in practice (Chen et al. 2018). Moreover, the surrogate endpoint for the interim decision should be carefully selected to be a good proxy for the primary endpoints.

Potential extensions and discussion to the flexible 2-in-1 design could be conducted in the future. Group sequential design method with the joint distribution of test statistics could be incorporated for the decision boundaries (Friede et al. 2020; Magnusson and Turnbull 2013; Wu et al. 2022). It could also be extended to multi-arm or multi-dose scenarios, to satisfy various needs and make good use of the seamless design. Moreover, the surrogate endpoint for interim decision could be also used in combination of the primary treatment effect for sample-size re-estimation via the calculation of modified conditional power (Li et al. 2021; Wu et al. 2022). Other multiplicity adjustment methods could be further explored for the type I error control.

**Disclosure statement**

The authors report there are no competing interests to declare.

**Supplementary materials**

Figure S1, Table S1-S2 are available online.

FDA (2014), "Guidance for Industry Expedited Programs for Serious Conditions – Drugs and Biologics."

FDA (2019), "Adaptive Designs for Clinical Trials of Drugs and Biologics - Guidance for Industry."

FDA (2020), "Accelerated Approval Program."

FDA (2022a), "Expansion Cohorts: Use in First-in-Human Clinical Trials to Expedite Development of Oncology Drugs and Biologics."

FDA (2022b), "Project FrontRunner -- Advancing Development of New Oncology Therapies to the Early Clinical Setting."

Fleming, T. R. (2006), "Standardversus adaptive monitoring procedures: a commentary," *Statistics in Medicine*, 25, 3305–3312. https://doi.org/10.1002/sim.2641.

Friede, T., Stallard, N., and Parsons, N. (2020), "Adaptive seamless clinical trials using early outcomes for treatment or subgroup selection: Methods, simulation model and their implementation in R," *Biometrical Journal*, 62, 1264–1283. https://doi.org/10.1002/bimj.201900020.

Gao, P., Ware, J. H., and Mehta, C. (2008), "Sample Size Re-Estimation for Adaptive Sequential Design in Clinical Trials," *Journal of Biopharmaceutical Statistics*, 18, 1184–1196. https://doi.org/10.1080/10543400802369053.

Li, Q., Lin, J., and Lin, Y. (2020), "Adaptive design implementation in confirmatory trials: methods, practical considerations and case studies," *Contemporary Clinical Trials*, 98, 106096. https://doi.org/10.1016/j.cct.2020.106096.

Li, Q., Lin, J., Liu, M., Wu, L., and Liu, Y. (2021), "Using Surrogate Endpoints in Adaptive Designs with Delayed Treatment Effect," *Statistics in Biopharmaceutical Research*, 1–10. https://doi.org/10.1080/19466315.2021.1938203.

Liu, A. (2022), "FDA oncology chief aims to open up accelerated approval for earlier cancer treatment under 'Project FrontRunner.'"

Magnusson, B. P., and Turnbull, B. W. (2013), "Group sequential enrichment design incorporating subgroup selection," *Statistics in Medicine*, 32, 2695–2714. https://doi.org/10.1002/sim.5738.

**Appendix**

**Solving Equation (4) by transforming semi-infinite integral to finite integral**

$$Pr(Y > Z_{1-\alpha}, X \leq c) + Pr(Z_2 > Z_{1-\alpha}, X > c) = \alpha$$

From Equation (6),

$$\Pr(Y > Z_{1-\alpha}, X \leq c) = \int_{-\infty}^{c} \Phi\left(\frac{\rho_{XY} x - Z_{1-\alpha}}{\sqrt{1 - \rho_{XY}^2}}\right) \phi(x) dx$$

From Equation (9),

$$Pr(Z_2 > Z_{1-\alpha}, X > c)$$

$$= \int_{W}^{\infty} \Pr(Z_2 > Z_{1-\alpha}|Z_1 = z_1) \Pr(X > c|Z_1 = z_1) f_{Z_1}(z_1) dz_1$$

$$+ \int_{-\infty}^{W} \Pr(Z_2 > Z_{1-\alpha}|Z_1 = z_1) \Pr(X > c|Z_1 = z_1) f_{Z_1}(z_1) dz_1$$

$$:= \int_{W}^{\infty} f_1(t) g(\rho_{XZ}, c) \phi(z_1) dz_1 + \int_{-\infty}^{W} f_2(n_2^*(t)) g(\rho_{XZ}, c) \phi(z_1) dz_1$$

Under the null hypothesis $H_0: \delta_X = 0$, $X|Z_1 = z_1 \sim N(\rho_{XZ} z_1, 1 - \rho_{XZ}^2)$,

$$g(\rho_{xz}, c) = \Pr(X > c|Z_1 = z_1) = \Phi\left(\frac{\rho_{XZ} z_1 - c}{\sqrt{1-\rho_{XZ}^2}}\right).$$

Let $\tilde{Z}_2$ denotes the incremental Z-test statistic based on the data accumulated after the interim analysis (i.e. based on $n_2^*$ subjects). Since $Z_2 = Z_1 \sqrt{n_1/n^*} + \tilde{Z}_2 \sqrt{n_2^*/n^*}$, and $n_2^* = \frac{n_1}{Z_1^2}\left(\frac{W - Z_1 t}{\sqrt{t(1-t)}}\right)^2$ and $t = n_1/n$,

$$f_1(t) = \Pr\left(\tilde{Z}_2 \geq \frac{Z_{1-\alpha}\sqrt{n} - Z_1\sqrt{n_1}}{\sqrt{n_2}} \Big| Z_1 = z_1\right)$$

$$= \Phi\left(\frac{z_1\sqrt{t} - Z_{1-\alpha}}{\sqrt{1 - t}}\right),$$

$$f_2(n_2^*(t)) = \Pr\left(\tilde{Z}_2 \geq \frac{Z_{1-\alpha}\sqrt{n^*} - Z_1\sqrt{n_1}}{\sqrt{n_2^*}} \middle| Z_1 = z_1\right) \Phi\left(\frac{z_1\sqrt{n_1} - Z_{1-\alpha}\sqrt{n}}{\sqrt{n_2}}\right)$$

$$= \Phi\left(\frac{z_1\sqrt{t} - Z_\alpha \sqrt{\min\left(t + \frac{t}{z_1^2}\left(\frac{W}{\sqrt{t(1-t)}} - z_1\sqrt{\frac{t}{1-t}}\right)^2, \frac{n_{max}}{n}\right)}}{\min\left(\left|\frac{\sqrt{t}}{z_1}\left(\frac{W}{\sqrt{t(1-t)}} - z_1\sqrt{\frac{t}{1-t}}\right)\right|, \sqrt{\frac{n_{max}}{n} - t}\right)}\right).$$

The integrals were transformed from infinite intervals to finite intervals by rules:

$$\int_a^\infty f(x)dx = \int_0^1 f\left(a + \frac{t}{1-t}\right)\frac{dt}{(1-t)^2} \text{ and } \int_{-\infty}^a f(x)dx = \int_0^1 f\left(a - \frac{1-t}{t}\right)\frac{dt}{t^2},$$

Then, the closed-form solution $C_{min}$ can be obtained by solving:

$$\alpha = \int_0^1 \Phi\left(\frac{\rho_{XY}\left(c - \frac{1-s}{s}\right) - Z_{1-\alpha}}{\sqrt{1-\rho_{XY}^2}}\right)\phi\left(c - \frac{1-s}{s}\right)\frac{ds}{s^2} +$$

$$\int_0^1 \Phi\left(\frac{(W + \frac{s}{1-s})\sqrt{t} - Z_{1-\alpha}}{\sqrt{1-t}}\right)\Phi\left(\frac{\rho_{XZ}(W + \frac{s}{1-s}) - c}{\sqrt{1-\rho_{XZ}^2}}\right)\phi\left(W + \frac{s}{1-s}\right)\frac{ds}{(1-s)^2} +$$

$$\int_0^1 \Phi\left(\frac{(W - \frac{1-s}{s})\sqrt{t} - Z_{1-\alpha}\sqrt{\min\left(t + \frac{t}{(W-\frac{1-s}{s})^2}\left(\frac{W}{\sqrt{t(1-t)}} - (W-\frac{1-s}{s})\sqrt{\frac{t}{1-t}}\right)^2, \frac{n_{max}}{n}\right)}}{\min\left(\left|\frac{\sqrt{t}}{(W-\frac{1-s}{s})}\left(\frac{W}{\sqrt{t(1-t)}} - (W-\frac{1-s}{s})\sqrt{\frac{t}{1-t}}\right)\right|, \sqrt{\frac{n_{max}}{n} - t}\right)}\right)\Phi\left(\frac{\rho_{XZ}(W - \frac{1-s}{s}) - c}{\sqrt{1-\rho_{XZ}^2}}\right)\phi\left(W - \frac{1-s}{s}\right)\frac{ds}{s^2}.$$

Supplementary Materials for:

**Flexible Seamless 2-in-1 Design with Sample Size Adaptation**

Figure S1. Simulated power of Chen's 2-in-1 design (S2in1) deisgn by different effect size (HR in 0.7-0.55) (N=100,000). Number of events at final analysis for phase 2 is 118, and 330 for phase 3. $X$ and $Y$ are drawn from standardized bivariate normal distribution with correlation $\rho_{XY}$, $X$ and $Z$ are drawn from standardized bivariate normal distribution with correlation $\rho_{XZ}$. $\alpha = 0.025$ (one-sided), $1 - \beta = 0.9$.

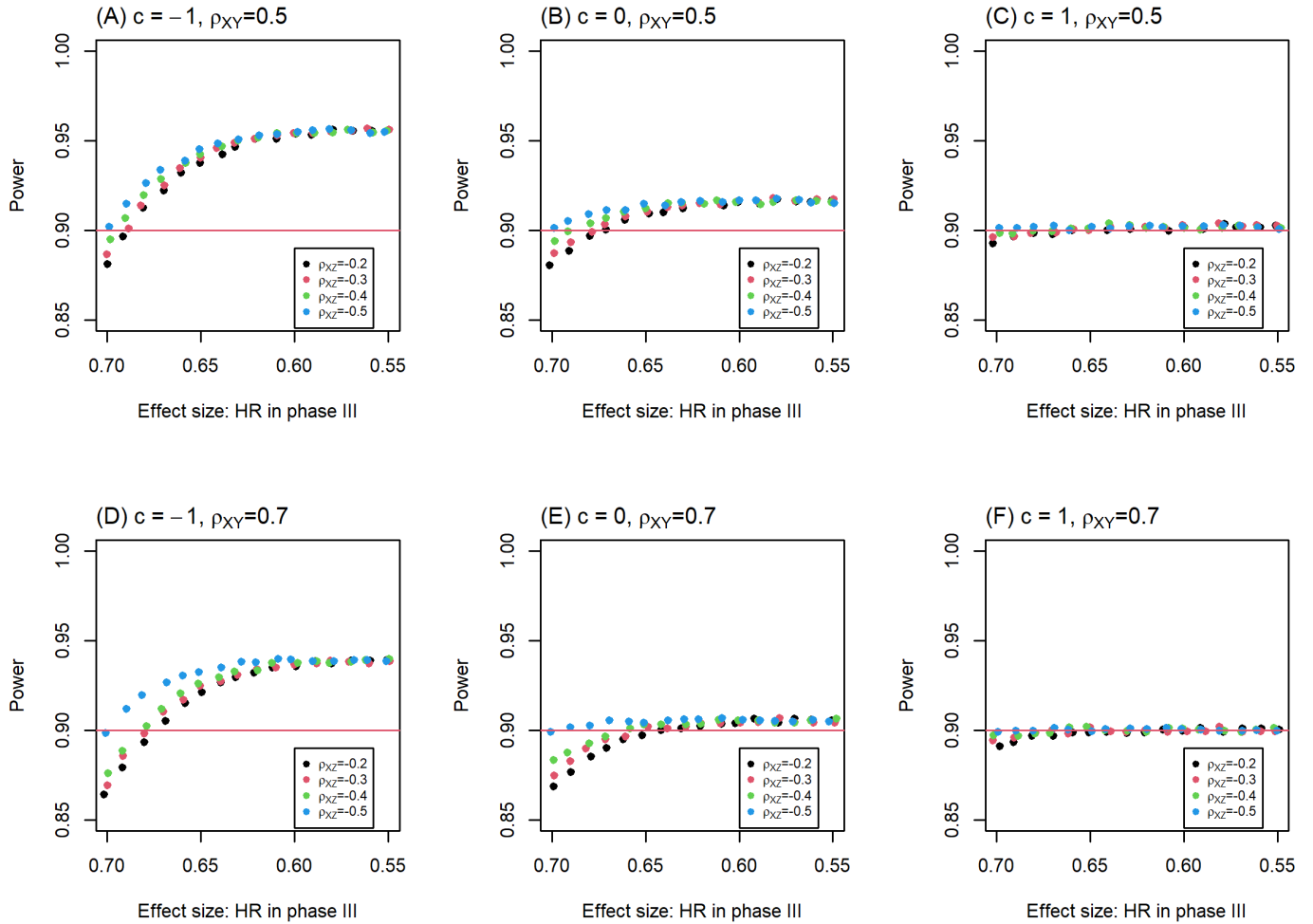

Table S1. Operating characteristics (number of Monte-Carlo iterations N=10,000) with different correlation between intermediate endpoint and primary endpoint in phase 3 $\rho_{XZ}$. $HR_{OS}=HR_{PFS}=0.55$, ORRs (ORR of control and treatment groups) = (0.1,0.3). Correlation between intermediate endpoint and primary endpoint in phase 2 is fixed as $\rho_{XY}=0.7$. Number of events before interim analysis $m_1=60$, number of events at final analysis for phase 3 in Chen's 2-in-1 design (S2in1) could be $m_1+m_2=180$ or $m_{max}=330$. Power of using CHW statistics for the flexible 2-in-1 design with sample size re-estimation (F2in1) is also calculated for phase 3.

| $\rho_{XZ}$ | Interim outcome | Probability of interim outcome | Power conditional on interim outcome | | | | Expected event size | | | Expected study duration | | |
|---|---|---|---|---|---|---|---|---|---|---|---|---|
| | | | S2in1 | | F2in1 | F2in1 w/ CHW | S2in1 | | F2in1 | S2in1 | | F2in1 |
| | | | $m_1+m_2$ | $m_{max}$ | | | $m_1+m_2$ | $m_{max}$ | | $m_1+m_2$ | $m_{max}$ | |
| 0.7 | Overall | - | 0.913 | 0.921 | 0.918 | - | 163 | 271 | 173 | 37 | 52 | 39 |
| | X<=c: Phase 2 | 27.8% | 0.715 | 0.715 | 0.715 | - | 118 | 118 | 118 | 30 | 30 | 30 |
| | X>c: Phase3 | 72.3% | 0.989 | 1.000 | 0.995 | 0.995 | 180 | 330 | 194 | 40 | 61 | 42 |
| 0.6 | Overall | - | 0.910 | 0.917 | 0.915 | - | 163 | 272 | 175 | 38 | 52 | 39 |
| | X<=c: Phase 2 | 27.6% | 0.700 | 0.700 | 0.700 | - | 118 | 118 | 118 | 30 | 30 | 30 |
| | X>c: Phase3 | 72.4% | 0.990 | 1.000 | 0.997 | 0.997 | 180 | 330 | 197 | 40 | 61 | 43 |
| 0.5 | Overall | - | 0.906 | 0.918 | 0.914 | - | 163 | 272 | 178 | 37 | 52 | 39 |
| | X<=c: Phase 2 | 27.6% | 0.704 | 0.704 | 0.704 | - | 118 | 118 | 118 | 29 | 29 | 29 |
| | X>c: Phase3 | 72.4% | 0.984 | 1.000 | 0.994 | 0.994 | 180 | 330 | 201 | 40 | 61 | 43 |
| 0.4 | Overall | - | 0.909 | 0.920 | 0.918 | - | 164 | 274 | 181 | 37 | 53 | 40 |
| | X<=c: Phase 2 | 26.4% | 0.698 | 0.698 | 0.698 | - | 118 | 118 | 118 | 29 | 29 | 29 |
| | X>c: Phase3 | 73.6% | 0.985 | 1.000 | 0.997 | 0.997 | 180 | 330 | 203 | 40 | 61 | 44 |

Table S2. Operating characteristics (number of Monte-Carlo iterations N=10,000) with different $m_{max}$. $HR_{OS}=HR_{PFS} = 0.55$, ORRs (ORR of control and treatment groups) = (0.1,0.3). Correlation between intermediate endpoint and primary endpoint in phase 2 is fixed as $\rho_{XY} = 0.7$. Number of events before interim analysis $m_1 = 60$, number of events at final analysis for phase 3 in Chen's 2-in-1 design (S2in1) could be $m_1 + m_2 = 180$ or $m_{max} = 330$. Power of using CHW statistics for the flexible 2-in-1 design with sample size re-estimation (F2in1) is also calculated for phase 3.

| $m_{max}$ | Interim outcome | Probability of interim outcome | Power conditional on interim outcome | | | | Expected event size | | | Expected study duration | | |
|---|---|---|---|---|---|---|---|---|---|---|---|---|
| | | | S2in1 | | F2in1 | F2in1 w/ CHW | S2in1 | | F2in1 | S2in1 | | F2in1 |
| | | | $m_1 + m_2$ | $m_{max}$ | | | $m_1 + m_2$ | $m_{max}$ | | $m_1 + m_2$ | $m_{max}$ | |
| 400 | Overall | - | 0.909 | 0.921 | 0.917 | - | 163 | 323 | 182 | 37 | 59 | 40 |
| | X<=c: Phase 2 | 27.3% | 0.709 | 0.709 | 0.709 | - | 118 | 118 | 118 | 30 | 30 | 30 |
| | X>c: Phase3 | 72.7% | 0.985 | 1.000 | 0.995 | 0.995 | 180 | 400 | 206 | 40 | 71 | 44 |
| 360 | Overall | - | 0.909 | 0.917 | 0.915 | - | 163 | 294 | 180 | 38 | 55 | 40 |
| | X<=c: Phase 2 | 27.3% | 0.696 | 0.696 | 0.696 | - | 118 | 118 | 118 | 30 | 30 | 30 |
| | X>c: Phase3 | 72.7% | 0.988 | 1.000 | 0.996 | 0.996 | 180 | 360 | 203 | 40 | 65 | 44 |
| 330 | Overall | - | 0.906 | 0.918 | 0.914 | - | 163 | 272 | 178 | 37 | 52 | 39 |
| | X<=c: Phase 2 | 27.6% | 0.704 | 0.704 | 0.704 | - | 118 | 118 | 118 | 29 | 29 | 29 |
| | X>c: Phase3 | 72.4% | 0.984 | 1.000 | 0.994 | 0.994 | 180 | 330 | 201 | 40 | 61 | 43 |
| 270 | Overall | - | 0.911 | 0.920 | 0.918 | - | 163 | 229 | 174 | 37 | 47 | 39 |
| | X<=c: Phase 2 | 26.7% | 0.703 | 0.703 | 0.703 | - | 118 | 118 | 118 | 29 | 29 | 29 |
| | X>c: Phase3 | 73.3% | 0.986 | 0.999 | 0.996 | 0.997 | 180 | 270 | 194 | 40 | 53 | 42 |